\documentclass{PoS}
\usepackage{amsmath,amssymb,bm,comment,color, graphicx}

\newcommand{\lbr}{\left(}
\newcommand{\rbr}{\right)}

\newcommand{\CS}{S}

\newcommand{\arXiv}[1]{{#1}}
\newcommand{\url}[1]{{#1}}
\newcommand{\texorpdfstring}[1]{{#1}}

\def\d{\partial}

\def\half{\frac{1}{2}}

\def\nn{\nonumber}

\title{Chiral magnetic effect and holography}

\ShortTitle{Chiral magnetic effect and holography}

\author{\speaker{Ingo Kirsch}
\\
         DESY Hamburg, Theory Group,  Notkestrasse 85, D-22607 Hamburg, Germany\\
        E-mail: \email{ingo.kirsch@desy.de}}

\author{Tigran Kalaydzhyan\\
          DESY Hamburg, Theory Group,  Notkestrasse 85, D-22607 Hamburg, Germany\\
       E-mail: \email{tigran.kalaydzhyan@desy.de}}

\abstract{The chiral magnetic effect (CME) is a highly discussed effect in heavy-ion collisions stating that, in the presence of a magnetic field $B$, an electric current is generated in the background of topologically nontrivial gluon fields. We present a holographic (AdS/CFT) description of the CME in terms of a fluid-gravity model which is dual to a strongly-coupled plasma with multiple anomalous U(1) currents. In the case of two U(1) charges, one axial and one vector, the CME formally appears as a first-order transport coefficient in the vector current. We will holographically compute this coefficient at strong coupling and compare it with the hydrodynamic result. Finally, we will discuss an anisotropic variant of the model and study a possible dependence of the CME on the elliptic flow coefficient $v_2$. \hfill {DESY 13-016}
}

\FullConference{Xth Quark Confinement and the Hadron Spectrum,\\
		October 8-12, 2012\\
		TUM Campus Garching, Munich, Germany}

\begin{document}

\section{Introduction}

The chiral magnetic effect (CME) is a hypothetical phenomenon which states
that, in the presence of a magnetic field $B$, a nonzero axial charge density will
lead to an electric current along the direction of the $B$ field \cite{Kharzeev}. While there is common agreement that
the chiral magnetic effect may occur in off-central heavy-ion collisions, it is still under
intense discussion whether or not current measurements of
charge-dependent correlation functions are sensitive enough to the CME, see \cite{Liao}
for a  phenomenological analysis. Regardless of how the
CME will eventually manifest itself in heavy-ion collisions, we discuss both a hydrodynamical
as well as a holographic (AdS/CFT) model for the CME. The fluid-gravity duality
 seems to be the right framework for a holographic description, as it captures
the response of the system to an external
perturbation, in case of the CME the perturbation by a $B$ field.
 In this talk we review the fluid-gravity
 model proposed in \cite{Kirsch11, Anisotropic} and derive the CME (and
related effects) from it.
Other holographic approaches to the CME can be found in
\cite{Lifschytz}--\cite{Landsteiner:2012kd}.

\section{Chiral Magnetic Effect in Hydrodynamics}

\subsection{Hydrodynamics with $n$ anomalous $U(1)$ charges}

The hydrodynamic regime of isotropic relativistic fluids with triangle
anomalies has been studied in \cite{Son}--\cite{Oz}.
Such
fluids typically contain $n$ anomalous $U(1)$ charges which commute
with each other.  The stress-energy
tensor $T^{\mu\nu}$ and $n$ $U(1)$ currents $j^{a\mu}$ ($a=1,...,n$)  are
\begin{align}
T^{\mu\nu}&= (\epsilon+P) u^\mu u^\nu + P g^{\mu\nu} + \tau^{\mu\nu} \,,
\qquad
j^{a\mu} = \rho^a u^\mu + \nu^{a \mu}\,, \label{currents}
\end{align}
where $\rho^a$, $\epsilon$, and $P$ denote the charge densities, energy density and pressure,
respectively.  $\tau^{\mu\nu}$ and $\nu^{a \mu}$ denote higher-gradient corrections, and
 $g_{\mu\nu}$ is the metric with signature
$(-,+,+,+)$.

 The corresponding hydrodynamic equations
are
\begin{align}\label{hydroeqn}
  \partial_\mu T^{\mu\nu} = F^{a\nu\lambda} j^{a}_\lambda
  \,,\qquad
  \d_\mu j^{a\mu} = C^{abc} E^b\cdot B^c \,,
\end{align}
where $E^{a\mu}=F^{a \mu\nu} u_\nu$, $B^{a\mu} =
\frac{1}{2}\epsilon^{\mu\nu\alpha \beta}u_\nu F^a_{\alpha\beta}$
are electric and magnetic fields, and
$F^a_{\mu\nu}=\partial_\mu A^a_\nu-\partial_\nu A^a_\mu$ are the
gauge field strengths.  The anomaly coefficients are given by a totally
symmetric \mbox{rank-$3$} tensor~$C^{abc}$.
 As in \cite{Son}, we expand the constitutive equations
for $T^{\mu\nu}$ and $j^\mu$ up to first order, taking  $A^a_\mu \sim O(p^0)$
and $F^a_{\mu\nu} \sim O(p)$. The gauge fields $A^a_\mu$ are non-dynamical.

The response of the system to the application of external magnetic fields and rotation
is measured by the chiral magnetic and vortical conductivities $\xi_B^{ab}$ and
$\xi^a_\omega$, respectively. The first-order correction of the $U(1)$
currents is given by
\begin{equation}
  \nu^{a\mu} = \xi^a_\omega \omega^\mu + \xi_B^{ab} B^{b\mu}
\,,
\end{equation}
where $\omega^\mu\equiv \frac{1}{2} \epsilon^{\mu\nu\lambda\rho} u_\nu\partial_\lambda u_\rho$ is the
vorticity.
The conductivities $\xi^a_\omega$ and $\xi_B^{ab}$ were first introduced
in \cite{Erdmenger, Son} and are given by \cite{Son, Oz}
\begin{align}
  \xi^a_\omega &= C^{abc} \mu^b\mu^c
    - \frac{2}{3} \rho^a C^{bcd} \frac{\mu^b\mu^c\mu^d}{\epsilon+P} + {\cal O}(T^2)   \,,\label{xi}\\
 \xi_B^{ab} &=  C^{abc} \mu^c
    - \frac{1}{2}\rho^a C^{bcd} \frac{\mu^c\mu^d}{\epsilon+P} + {\cal O}(T^2)   \,,\label{xiB}
\end{align}
where $\mu^a$ are the corresponding chemical potentials. The terms in ${\cal O}(T^2)$
are related to gravitational anomalies~\cite{Landsteiner:2011cp}, which we do not discuss here.

\subsection{Chiral magnetic and vortical effect \texorpdfstring{$(n=2)$}{}}\label{sec2D}

Physically, the most interesting case is that involving two charges
($n=2$) \cite{Isachenkov,Pu, Kirsch11}. The chiral magnetic effect \cite{Kharzeev} can be described
by one axial and one vector $U(1)$, denoted by $U(1)_A \times U(1)_V$.
A convenient notation for the gauge fields and currents is
($a,b,...=1,2$)
\begin{align}
{A}_\mu^A &= {A}^1_\mu \,,\qquad {A}_\mu^V = {A}^2_\mu \,,\nonumber\\
j^{\mu}_5&=j^{1\mu}  \,, \qquad j^{\mu}= j^{2\mu} \,.
\end{align}

Let us now derive the chiral magnetic and vortical effects from
(\ref{xi}) and (\ref{xiB}).  $C-$parity allows for $C^{111} \neq 0$
and $C^{122}=C^{221}=C^{212}\neq 0$, while $C^{121}=C^{211}=C^{112}=C^{222}=0$~\cite{Anisotropic}.
The hydrodynamic equations (\ref{hydroeqn}) then imply non-conserved
vector and axial currents,
\begin{align}
 \partial_\mu j^\mu &= \textstyle -\frac{1}{4} (C^{212} F^A_{\mu\nu} \tilde F^{V\mu\nu}
 + C^{221} F^V_{\mu\nu} \tilde F^{A\mu\nu}) \,, \nonumber\\
 \partial_\mu j^\mu_5 &=  \textstyle -\frac{1}{4}(C^{111} F^A_{\mu\nu} \tilde F^{A\mu\nu}
+ C^{122} F^V_{\mu\nu}\tilde F^{V\mu\nu}) \,,
\end{align}
where we rewrote $E^b \cdot B^c = -\frac{1}{4}F^b_{\mu\nu}\tilde
F^{c\,\mu\nu}$ (with $\tilde F^{a \mu\nu}=\frac{1}{2}\varepsilon^{\mu\nu\rho\sigma}F^a_{\rho\sigma}$).

To restore the conservation of the vector current, we add
the Bardeen currents
\begin{align}
  j^\mu_B &= c_B \varepsilon^{\mu\nu\lambda\rho} (A^V_\nu
  F_{\lambda\rho}^A
  - 2 A_\nu^A F^V_{\lambda\rho} )\,,
\qquad
  j^\mu_{5,B} = c_B \varepsilon^{\mu\nu\lambda\rho} A^V_\nu
  F^V_{\lambda\rho} \,,\qquad (c_B=-C^{122}/2)
\label{Bardeen}
\end{align}
 to the vector and axial currents,
\begin{align}
  &j'{}^{\mu} \equiv j^\mu + j_B^\mu\,,\qquad j'{}^{\mu}_{\!\!5\,\,}
  \equiv j^\mu_5 + j_{5,B}^\mu\,.
\end{align}
Setting also $C^{111}=C^{122} \equiv C/3$, the hydrodynamic equations
(\ref{hydroeqn}) become
\begin{align}
  \partial_\mu T^{\mu\nu} &=  F^{V \nu\lambda} j'_{\lambda}
  +  F^{A \nu\lambda} j'_{5\lambda} \,, \nonumber \\
  \partial_\mu j'{}^{\mu} &= 0 \,,\nonumber\\
  \partial_\mu j'{}^{\mu}_{\!\!5\,\,} &= C  E \cdot  B + (C/3 )E_5 \cdot  B_5  \,.
\end{align}

Using the derivative expansion
\begin{align}
  j'{}^{\mu} & = \rho u^\mu + \kappa_\omega\omega^\mu + \kappa_B
  B^{\mu} + \kappa_{5,B} B_5^{\mu} \,,
\end{align}
where $\kappa_\omega \equiv \xi^2_\omega$, $\kappa_B \equiv
\xi_B^{22}$ and $\kappa_{5,B} \equiv \xi_B^{21}$, we obtain from
(\ref{xi}) and (\ref{xiB}) the conductivities\footnote{$\kappa_{5,B}$ represents
another effect, which occurs only in the presence of an axial $B$ field~\cite{Anisotropic}.}
\begin{align}
  \kappa_\omega &= 2 C \mu_5 \left( \mu - \displaystyle\frac{\rho}{\epsilon + P} \left[\mu^2+\frac{\mu^2_5}{3} \right] \right),\qquad
  \quad \kappa_B = C \mu_5 \left( 1 - \displaystyle\frac{\mu \rho}{\epsilon + P} \right) .
 \label{kappa}
\end{align}
where $\mu_5\equiv \mu^1$, $\mu\equiv\mu^2$.
There are analogous transport coefficients in the axial current
$j^\mu_5$ \cite{Kirsch11}.  The axial fields $E_{5\mu}$ and $B_{5\mu}$
are not needed and can now be switched off.
The first term in $\kappa_B$ and $\kappa_\omega$, $\kappa_B =
C\mu_5$ and $\kappa_\omega=2C\mu\mu_5$, is the leading term in the {\em chiral
  magnetic} (CME) \cite{Kharzeev} and {\em chiral vortical
  effect} (CVE) \cite{CVE}, respectively.
The second term proportional to  $\rho/(\epsilon+P)$
actually depends on the dynamics of the fluid.\footnote{In \cite{Zakharov}
this term was considered as a one-loop correction in an effective theory and $(\epsilon+P)/\rho$
was interpreted as the corresponding  infrared cutoff in the energy/momentum
integration.}

\section{Fluid-gravity model for the chiral magnetic effect}\label{sec3}

In this section we construct the gravity dual of an isotropic fluid
with $n$ anomalous $U(1)$ charges.
We start from a five-dimensional $U(1)^n$ Einstein-Maxwell theory
in an asymptotic AdS space. The action is
\begin{align}
  S&=\frac{1}{16\pi G_5} \int d^5x \sqrt{-g} \left[ R - 2\Lambda
    - F^a_{MN} F^{aMN}  + \frac{S_{abc}}{6 \sqrt{-g}}
    \varepsilon^{PKLMN} A^a_P F^b_{KL} F^c_{MN} \right] \,, \nonumber
\end{align}
where $\Lambda=-6$ is the cosmological constant. As usual, the $U(1)$
field strengths are defined by
$ F^a_{MN} = \partial_M A^a_N - \partial_N A^a_M$,
where $M,N,... = 0,...,4$ and $a=1,...,n$. The Chern-Simons term
$A\wedge F \wedge F$ encodes the information of the triangle anomalies
in the field theory \cite{Son}. In fact, the Chern-Simons coefficients
$S_{abc}$ are related to the anomaly coefficients $C_{abc}$ by
\begin{align}
  C_{abc} = S_{abc}/(4\pi G_5) \label{CSrel} \,.
\end{align}

The corresponding equations of motion are given by the combined system
of Einstein-Maxwell and Maxwell equations,
\begin{align}
  G_{MN} - 6 g_{MN} &= T_{MN}
  \,, 
\qquad
  \nabla_M F^{aMP} = -\frac{S_{abc}}{8\sqrt{-g}} \varepsilon^{PMNKL}
  F^b_{MN} F^c_{KL} \,,
\label{eom}
\end{align}
where the energy-momentum tensor $T_{MN}$ is
\begin{align}
  T_{MN} &= - 2 \left( F^a_{MR} F^{aR}{}_N + \frac{1}{4} g_{MN}
    F^a_{SR} F^{aSR} \right) \,.
\end{align}

\subsection{AdS black hole with multiple U(1) charges}

A gravity dual to an {\em isotropic} fluid ($\epsilon=3P$) with
multiple chemical potentials $\mu^a$ ($a=1,...,n$) at finite
temperature $T$ is given by an AdS black hole solution with mass $m$
and multiple $U(1)$ charges $q^a$.  In Eddington-Finkelstein
coordinates, the metric and $U(1)$ gauge fields  are
\begin{align}
  ds^2 &= - f(r) dt^2 + 2 dr dt + r^2 d\vec x^2\,,
\qquad
  A^a  = - A^a_0(r) dt \,, \label{ansatz0th}
\end{align}
where
\begin{align}
  f(r)&= r^2 - \frac{m}{r^2} +\sum_a \frac{(q^a)^2}{r^4}\,,\qquad
  A^a_0(r) = \mu^a_\infty + \frac{\sqrt{3} q^a}{2 r^2} \,.
\label{param}
\end{align}
In case of a single charge ($n=1$), the
background reduces to an ordinary Reissner-Nordstr\o m black hole
solution in $AdS_5$ \cite{Chamblin}.

The temperature $T$ and chemical potentials $\mu^a$ of the fluid are
defined by
\begin{align}
  T &
= \frac{f'(r_+)}{4\pi} =
      \frac{2r_+^6 -  \sum_a (q_a)^2}{2\pi r_+^5} \,,
\qquad
  \mu^a = A^a_0(r_\infty)- A^a_0(r_+)   \,,
\end{align}
where $r_+$ is the outer horizon defined by the maximal solution of
$f(r)=0$, and the boundary is located at $r_\infty$.  The
temperature of the fluid is the Hawking temperature of the black hole.

\section{Holographic vortical and magnetic conductivities}\label{sec4}

We will now compute the chiral vortical and magnetic conductivities $\xi^a_\omega$ and
$\xi^{ab}_B$ from first-order corrections to the AdS
geometry (\ref{ansatz0th}) using the fluid-gravity correspondence \cite{Hubeny}.

\subsection{First-order corrected background and chiral conductivities}

In order to become a dual to a multiply-charged fluid, the AdS geometry
(\ref{ansatz0th}) must be boosted along the four-velocity of the fluid
$u_\mu$ ($\mu=0,...,3$). The boosted version of (\ref{ansatz0th}) is
\begin{align}
  ds^2 &= \left(r^2  P_{\mu\nu}-f(r) u_\mu u_\nu\right) dx^\mu dx^\nu -2u_\mu dx^\mu dr  \,,\qquad
  A^a = (A_0^a(r) u_\mu + {\cal A}^a_\mu ) dx^\mu \,,
\label{0thordersol}
\end{align}
where $P^{\mu\nu} = g^{\mu\nu} + u^\mu u^\nu$, and $f(r)$ and $A_0^a(r)$ as in (\ref{param}).
Following
\cite{Son, Kirsch11}, we have formally introduced constant background
gauge fields ${\cal A}^a_\mu$ to model external electromagnetic
fields, such as the magnetic fields $B^{a\mu}$ needed for the chiral
magnetic effect.

The transport coefficients $\xi^a_\omega$ and $\xi^{ab}_B$ can now be
computed using standard fluid-gravity techniques \cite{Hubeny}. We
closely follow \cite{Son, Yee, Kirsch11}, in which these transport
coefficients were determined for an isotropic fluid with one and three
charges ($n=1,3$).  We work in the static frame $u_\mu=(-1,0,0,0)$
and consider vanishing background fields ${\cal
  A}^a_\mu$ (at $x^\mu=0$). The transport coefficients $\xi^a_\omega$
and $\xi_B^{ab}$ measure the response of the system to rotation and
the perturbation by an external magnetic field.  We therefore slowly
vary $u_{\mu}$ and
${\cal A}^a_\mu$ up to first order as ($i=1,2,3$)
\begin{align}
  u_{\mu} = (-1, x^{\nu}\partial_{\nu}u_i)\,,\qquad {\cal A}^a_\mu =
  (0, x^{\nu}\partial_{\nu} {\cal A}^a_i) \,.
\label{variations}
\end{align}
Due to the dependence on $x^\mu$, the background (\ref{0thordersol})
is no longer an exact solution of the equations of motion. Instead
with varying para\-meters the solution (\ref{0thordersol}) receives
higher-order corrections, which are in this case of first order in the
derivatives.

An ansatz for the first-order corrected metric and gauge fields is
given by
\begin{align}
  ds^2 &= \lbr -f(r)+\tilde g_{tt}\rbr dt^2 +2 \lbr 1+\tilde g_{tr} \rbr dt dr
+ r^2   (   dx^2 +  dy^2 + dz^2)
\nonumber\\
 &~~~+\tilde g_{ij}dx^i dx^j-2 x^{\nu} \partial_\nu u_i dr dx^i
+ 2\lbr \lbr f(r) - r^2 \rbr x^\nu\partial_\nu u_i +
  \tilde g_{ti}\rbr dt dx^i \,, \nn\\
  A^a&=\lbr -A^a_0(r) +\tilde A^a_t\rbr dt
+ \lbr A^a_0(r) x^\nu \partial_\nu u_i + x^\nu \partial_\nu{\cal
    A}^a_i + \tilde A^a_i \rbr dx^i\,, \label{ansatz}
\end{align}
where we denote the first-order corrections by
  $\tilde g_{MN} = \tilde g_{MN}(r)$
and $\tilde A_M^a = \tilde   A_M^a(r)$
and choose the gauge
  $\tilde g_{rr}=0$, $\tilde g_{r\mu}\sim u_\mu$, $\tilde
  A^a_r=0$, $\sum_{i=1}^{3}\tilde g_{ii}=0$.

The first-order corrections can be obtained by substituting the ansatz
(\ref{ansatz}) into the equations of motion (\ref{eom}).
The computation is straight-forward but lengthy and
can be found in appendix~C of \cite{Anisotropic} (To obtain the corrections for an isotropic
background, set $w_L=w_T=1$ there \cite{Anisotropic}).
For the magnetic and vortical
effects, we only need the gauge field corrections $\tilde A^a_\mu(r)$ given by
\begin{align}
 \tilde A^a_t = 0 \,,\qquad
  \tilde A^a_i(r) &= \int_\infty^r dr' { \frac{1}{ r' f(r')} } \left[ Q_i^a(r') - Q_i^a(r_+)
-  C_i {r_+} A_0^a{}'(r_+) + {r' \tilde
      g_{ti}(r')} A_0^a{}'(r') \right] \,,
\end{align}
where $Q_i^a$ and $C_i$ are defined by
\begin{allowdisplaybreaks}
\begin{align}
  Q_i^a &\equiv\textstyle \frac{1}{2}\CS^{abc} A_0^b A_0^c \epsilon^{ijk}\left(
\partial_j u_k\right)+ \CS^{abc} A_0^b \epsilon^{ijk}\left(\partial_j {\cal A}^c_k\right), \nn \\
C_i &=  4 c(r_+)
 \Big(\textstyle\frac{1}{3} \CS^{abc}{A_0^a(r_+)A_0^b(r_+)A_0^c(r_+)}\epsilon^{ijk}\left(\partial_j u_k\right)
+\textstyle \frac{1}{2} \CS^{abc}{A_0^a(r_+)A_0^b(r_+) }\epsilon^{ijk}\left(\partial_j {\cal A}^c_k\right)\Big)\,, \nonumber\\
&~~~~~~c(r_+) =\left[ { { {r_+ (f'(r_+)- {4}
\sum_{a}  A_0^a(r_+) A_0^a{}'(r_+)) }}}\right]^{-1}\,.\nonumber
\end{align}
\end{allowdisplaybreaks}
The term involving  $\tilde g_{ti}(r')$ will not be needed in the following.


On the boundary of the asymptotic AdS space (\ref{ansatz}), the metric
and gauge fields couple to the fluid stress-energy tensor and $U(1)$
currents, respectively. Holographic renormalization \cite{Skenderis}
provides relations between these currents and the near-boundary
behavior of their dual bulk fields.  The $U(1)$ currents $j^{a\mu}$ are
related to the bulk gauge fields $A^{a\mu}$ by \cite{Skenderis, Yee3}
\begin{align}
  j^{a \mu} &= \frac{1}{2\pi
    G_5} \left[ \eta^{\mu\nu} A^{a(2)}_\nu - {\hat j}^{a\mu}\right] \,,\qquad
\hat j^{a\mu} \propto {\CS^{abc}}
  \epsilon^{\mu\nu\rho\sigma}A_{\nu}^{b(0)}\partial_\rho A_{
    \sigma}^{c(0)}  \,, \label{hr}
\end{align}
where $A_{\mu}^{a(n)}$ are the $n$th-order coefficients in a
$\frac{1}{r}$-expansion of the bulk
gauge fields $A^a_{\mu}(r, x^\mu)$.  Setting ${A}_\nu^{a(0)}= \mu^a_\infty u_\nu=0$ allows
us to ignore contributions from $\hat j^{a\mu}$.

Expanding the solution in $\frac{1}{r}$ and substituting only the
corrections $\tilde A^{a}_{\mu}$, we get the currents
\begin{align}
  \tilde j^{a \mu} &= \lim_{r\rightarrow\infty}\frac{r^2}{2\pi G_5}\eta^{\mu\nu}\tilde A^a_\nu(r)
  = \frac{1}{4\pi G_5} \eta^{\mu\nu} \lbr Q_\nu^{a}(r_+) +{r_+ }
  A_0^a{}'(r_+) C_\nu \rbr\,.\label{currentgravity}
\end{align}
Substituting (\ref{param}) into the right-hand side of  (\ref{currentgravity})
and comparing the result with\footnote{ Use also
$4  r_+ A_0^{a\prime}(r_+) c(r_+) = {\sqrt{3} q^a }/{m}= {\rho^a}/{(\varepsilon+P)}$
for the prefactor of the second term in  (\ref{currentgravity})  \cite{Anisotropic} (We thereby
correct a factor 4 in \cite{Anisotropic}. Note that $\rho^a= j^{a0}=\sqrt{3}q^a/(4\pi G_5)$).
}
\begin{align}
  \tilde j^{a \mu}&= \xi^a_\omega\, \omega^\mu + \xi_B^{a b}\,B^{b\mu}
  = \xi^a_\omega \,\textstyle
  \half\epsilon^{\nu\rho\sigma\mu}u_\nu\partial_\rho u_\sigma +
  \xi_B^{a b} \epsilon^{\nu\rho\sigma\mu}u_\nu\partial_\rho {\cal
    A}^b_\sigma\, ,  \label{current_hydrodynamics}
\end{align}
we finally obtain the coefficients
\begin{align}
  \xi^a_\omega&= \frac{1}{4 \pi G_5}\Bigg( \CS^{abc} {\mu^b\mu^c}
  - \frac{2}{3} \frac{\rho^a}{\epsilon + P} \CS^{bcd} \mu^b\mu^c\mu^d \Bigg)
  \label{xi_hol}\,,\\
  \xi^{ab}_B &= \frac{1}{ 4 \pi G_5}\Bigg( \CS^{abc} {\mu^c} -
  \frac{1}{2}\frac{\rho^a}{\epsilon + P} \CS^{bcd} \mu^c\mu^d
  \Bigg)\,,\label{xiB_hol}
\end{align}
with $\mu^a\equiv - A^a_0(r_+)$ (since $A^a_0(r_\infty)=\mu^a_\infty=0$),
in  agreement with  (\ref{xi}) and (\ref{xiB}). Restricting to two charges ($n=2$)
and performing similar manipulations as in section~\ref{sec2D}, we recover the
chiral conductivities $\kappa_\omega$ and $\kappa_B$  given by (\ref{kappa}).

\subsection{Subtleties in holographic descriptions of the CME}

The conservation of the electromagnetic current requires the
introduction of the Bardeen counter\-term into the action.  In AdS/QCD
models of the CME, this typically leads to a vanishing result for the
electromagnetic current \cite{Rubakov, Rebhan}. The problem is related
to the difficulty of introducing a chemical potential conjugated to a
non-conserved chiral charge \cite{Rubakov, Landsteiner}.  It is
possible to modify the action to obtain a conserved chiral
charge~\cite{Rubakov}.  This charge is however only gauge invariant
when integrated over all space in homogeneous configurations.

In AdS black hole models of the CME, one usually introduces a chiral
chemical potential dual to a gauge invariant current, despite it being
anomalous \cite{Landsteiner, Kirsch11}.
Consider the physical case with
two charges ($n=2$) as in section~\ref{sec2D} and define
axial and vector gauge fields by $A_{\mu}^{A}=A_{\mu}^{1}$ and
$A_{\mu}^{V}=A_{\mu}^{2}$.  Then $\hat j^\mu = \hat j_2^\mu$ gives
rise to an additional contribution of the type
\begin{align}
\hat j^{\mu}  \supset \varepsilon^{\mu\nu\rho\sigma}
{A}_\nu^{A(0)}(x) {F}^{V(0)}_{\rho\sigma}(x) \,,
\end{align}
which is however forbidden by electromagnetic gauge invariance
\cite{Rubakov}. This forces us to set ${A}_\nu^{A(0)}(x)= \mu_5^\infty u_\nu=0$ (at \mbox{$x=0$}).
 However, the chiral chemical potential $\mu_5=A^A_0(r_\infty)-A^A_0(r_+)= \mu_5^\infty - A^A_0(r_+) $
should be non-vanishing. Since  $\mu_5^\infty=0$, $A^A_0(r_+)$ must be non-vanishing, $A^A_0(r_+)\neq 0$, corresponding to
a non-vanishing gauge field at the horizon. In a static charged AdS black hole solution, one should set $A^A_0(r_+)= 0$
in order to get a well-defined gauge field at the horizon \cite{Chamblin}. However,
this is not necessarily required in the fluid-gravity duality, see \cite{Erdmenger},
in particular footnote~5 therein, and  $A^A_0(r_+) \neq 0$ does not lead to an inconsistency.

\section{CME in anisotropic fluids}

The above analysis can be repeated for anisotropic fluids with (rest-frame) stress-energy tensor
$T^{\mu\nu}={\rm diag}(\varepsilon, P_T, P_T, P_L)$ and different pressures
$P_T \neq P_L$, see \cite{Anisotropic} for details. It is expected that in this case  the
chiral magnetic conductivity $\kappa_B$ may depend on the anisotropy
parameter $\varepsilon_p=\frac{ P_T -P_L}{P_T+P_L}$. In fact, for small anisotropies
$\varepsilon_p$, we found (average pressure $\bar P=(2P_T+P_L)/3$) \cite{Anisotropic} 
\begin{align}\label{result}
 \kappa_B & \approx C \mu_5 \left( 1 - \displaystyle\frac{\mu \rho}{\epsilon + \bar P} \left[ 1-\frac{\varepsilon_p}{6}\right] \right) \,.
\end{align}
Despite the crudeness of the model, one can assume that such an anisotropic fluid describes to some extent
the anisotropic quark-gluon plasma, 
with our $\varepsilon_p$ imitating the real $\varepsilon_p \approx 2 v_2$ for pions.
Since the net chemical potential $\mu$ is quite small in current heavy-ion experiments,
the dependence on $\varepsilon_P$ (and hence $v_2$) in (\ref{result}) appears to be very mild. 
Even though the anisotropy dependence of $\kappa_B$ is very weak, 
(\ref{result}) tells us how the CME, if present in the experimental data, can be separated from the $v_2$-dependent background (for one of the 
attempts of such a separation see \cite{Liao}).

\end{document}